\documentclass[letterpaper,12pt]{JHEP3}
\usepackage{amsmath,amssymb}
\usepackage{latexsym}
\usepackage{epsfig}
\usepackage{cite}
\usepackage[english]{babel}

\newcommand{\eq}{\begin{equation}}
\newcommand{\feq}{\end{equation}}
\newcommand{\eqn}{\begin{eqnarray}}
\newcommand{\feqn}{\end{eqnarray}}
\newcommand{\arr}{\begin{eqnarray*}}
\newcommand{\farr}{\end{eqnarray*}}

\font\mybb=msbm10 at 12pt
\def\bb#1{\hbox{\mybb#1}}

\def\bR {\bb{R}}

\def\bC {\bb{C}}

\title{Rotating BPS black holes in matter-coupled AdS$_4$ supergravity}

\author{Dietmar Klemm$^{ab}$ \\
$^a$ Dipartimento di Fisica dell'Universit\`a di Milano, \\
\hspace*{0.15cm} Via Celoria 16, I-20133 Milano. \\
$^b$ INFN, Sezione di Milano, Via Celoria 16, I-20133 Milano. \\
}
\preprint{IFUM-975-FT}

\abstract{Using the general recipe given in arXiv:0804.0009, where all
timelike supersymmetric solutions of ${\cal N}=2$, $D=4$ gauged supergravity
coupled to abelian vector multiplets were classified, we construct
genuine rotating supersymmetric black holes in AdS$_4$
with nonconstant scalar fields. This is done for the $\text{SU}(1,1)/\text{U}(1)$ model
with prepotential $F=-iX^0X^1$. In the static case, the black holes are uplifted to
eleven dimensions, and generalize the solution found in hep-th/0105250
corresponding to membranes wrapping holomorphic curves in a Calabi-Yau
five-fold. The constructed rotating black holes preserve one quarter of the
supersymmetry, whereas their near-horizon geometry is one half BPS.
Moreover, for constant scalars, we generalize (a supersymmetric subclass of) the
Plebanski-Demianski solution of cosmological Einstein-Maxwell theory to an arbitrary
number of vector multiplets. Remarkably, the latter turns out to be related to the
dimensionally reduced gravitational Chern-Simons action.
}


\begin{document}

\section{Introduction}
\label{intro}

Black holes in anti-de~Sitter (AdS) spaces provide an important testground
to address fundamental questions of quantum gravity like holography. These
ideas originally emerged from string theory, but became
then interesting in their own right, for instance in recent applications to condensed
matter physics (cf.~\cite{Hartnoll:2009sz} for a review), where black holes
are again instrumental, since they provide the dual description of certain condensed
matter systems at finite temperature, like e.g.~holographic
superconductors \cite{Hartnoll:2008vx}.

On the other hand, among the extremal black holes (which have zero Hawking
temperature), those preserving a sufficient amount of supersymmetry are of
particular interest, as this allows (owing to non-renormalization theorems) to
extrapolate an entropy computation at weak string coupling (when the system
is generically described by a bound state of strings and branes) to the
strong-coupling regime, where a description in terms of a black hole is
valid \cite{Strominger:1996sh}. However, this picture, which has been essential
for our current understanding of black hole microstates, might be modified in
gauged supergravity (arising from flux compactifications) due to the presence of
a potential for the moduli, generated by the fluxes. This could even lead to a
stabilization of the dilaton, so that one cannot extrapolate between weak and
strong coupling anymore. Obviously, the explicit knowledge of supersymmetric
black hole solutions in AdS is a necessary ingredient if one wants to study this
new scenario.

A first step in this direction was made in \cite{Cacciatori:2009iz}, where the first
examples of extremal static BPS black holes in AdS$_4$ with nontrivial scalar
field profiles were constructed.
This analysis was facilitated by the results of \cite{Cacciatori:2008ek}, where all
timelike supersymmetric backgrounds of ${\cal N}=2$, $D=4$ gauged
supergravity coupled to abelian vector multiplets were classified. This provides
a systematic method to obtain BPS solutions, without the necessity to
guess some suitable ans\"atze. The upshot of \cite{Cacciatori:2009iz} was the
construction of a genuine static supersymmetric black hole with spherical
horizon. This came as a surprise, since up to now the
common folklore was that static spherical AdS black holes develop naked singularities
in the BPS limit \cite{Romans:1991nq}. This is indeed true in minimal
gauged supergravity, but the no-go theorems of \cite{Romans:1991nq}
were circumvented in \cite{Cacciatori:2009iz} by admitting
nonconstant moduli. The spherical solutions of \cite{Cacciatori:2009iz} were then
further studied and generalized in \cite{Hristov:2010ri,Dall'Agata:2010gj}.

In this paper, we shall go one step further with respect to \cite{Cacciatori:2009iz},
and include also rotation. Apart from the supersymmetric Kerr-Newman-AdS
family \cite{Kostelecky:1995ei,Caldarelli:1998hg} and its cousins with noncompact
horizons \cite{Caldarelli:1998hg},
there are not many known solutions of this type. One of the most notable exceptions
is perhaps the rotating two-charge black hole in $\text{SO}(4)$-gauged ${\cal N}=4$,
$D=4$ supergravity \cite{Chong:2004na}, whose BPS limit was studied in \cite{Cvetic:2005zi}.
Notice that the black holes constructed below are qualitatively different from the ones in
\cite{Chong:2004na}, since they are solitonic objects that admit no smooth limit
when the gauging is turned off.

In addition to the motivation given above, a further reason for considering
supersymmetric rotating black holes is the attractor mechanism \cite{Ferrara:1995ih,
Strominger:1996kf,Ferrara:1996dd,Ferrara:1996um,Ferrara:1997tw}, which states
that the scalar fields on the horizon and the entropy are independent of the asymptotic
values of the moduli. (The scalars are attracted towards their purely charge-dependent
horizon values). However, in gauged supergravity, the moduli fields have a potential, and
typically approach the critical points of this potential asymptotically, where
the solution approaches AdS. Thus, unless there are flat directions in the
scalar potential, the values of the moduli at infinity are completely fixed (in
terms of the gauge coupling constants), and therefore a more suitable formulation
of the attractor mechanism in AdS would be to say that the black hole entropy is
determined entirely by the charges, and is independent of the values of the moduli on
the horizon that are not fixed by the charges. First steps towards a systematic analysis of
the attractor flow in AdS were made in \cite{Morales:2006gm,Bellucci:2008cb} for the
non-BPS and in \cite{Cacciatori:2009iz,Dall'Agata:2010gj} for the BPS case, but it would
be very interesting to generalize in particular the results of \cite{Cacciatori:2009iz} to
include also rotation.

The remainder of this paper is organized as follows: In the next section, we
briefly review ${\cal N}=2$, $D=4$ gauged supergravity coupled to abelian vector
multiplets (presence of U$(1)$ Fayet-Iliopoulos terms) and give the general recipe
to construct supersymmetric solutions found in \cite{Cacciatori:2008ek}. In \ref{const-scal},
the equations of \cite{Cacciatori:2008ek} are solved for constant scalars.
This leads to a generalization of the Plebanski-Demianski solution
of cosmological Einstein-Maxwell theory to an arbitrary number of vector multiplets.
We also find a remarkable relationship of the latter with the dimensionally
reduced gravitational Chern-Simons action. In section \ref{nonconst-scal}, the case of
nonconstant scalars is considered, using the $\text{SU}(1,1)/\text{U}(1)$ model
with prepotential $F=-iX^0X^1$. First, we present in \ref{1/2BPS} a class of
one half BPS near-horizon geometries, where the moduli field still has a nontrivial
dependence on one of the horizon coordinates. Then, in section \ref{susy-rot-bh}, a
two-parameter family of rotating black holes is constructed. These solutions
preserve one quarter of the supersymmetries, and approach the geometries of
section \ref{1/2BPS} near the horizon. \eqref{lifting} contains an uplifting of
the obtained black holes to M-theory, together with some comments on their
higher-dimensional interpretation. We conclude in \ref{final} with some final remarks.

The reader who wants to skip the technical details can, instead of reading sections
\ref{const-scal} and \ref{susy-rot-bh}, immediately jump to eqns.~\eqref{PD-const}
ff.~and \eqref{metr-PD-scal} ff.~respectively for a summary of the results.

\section{${\cal N}=2$, $D=4$ gauged supergravity and its BPS geometries}
\label{sugra}

We consider ${\cal N}=2$, $D=4$ gauged supergravity coupled to $n_V$ abelian
vector multiplets \cite{Andrianopoli:1996cm}\footnote{Throughout this paper,
we use the notations and conventions of \cite{Vambroes}.}.
Apart from the vierbein $e^a_{\mu}$, the bosonic field content includes the
vectors $A^I_{\mu}$ enumerated by $I=0,\ldots,n_V$, and the complex scalars
$z^{\alpha}$ where $\alpha=1,\ldots,n_V$. These scalars parametrize
a special K\"ahler manifold, i.~e.~, an $n_V$-dimensional
Hodge-K\"ahler manifold that is the base of a symplectic bundle, with the
covariantly holomorphic sections
\begin{equation}
{\cal V} = \left(\begin{array}{c} X^I \\ F_I\end{array}\right)\,, \qquad
{\cal D}_{\bar\alpha}{\cal V} = \partial_{\bar\alpha}{\cal V}-\frac 12
(\partial_{\bar\alpha}{\cal K}){\cal V}=0\,, \label{sympl-vec}
\end{equation}
where ${\cal K}$ is the K\"ahler potential and ${\cal D}$ denotes the
K\"ahler-covariant derivative. ${\cal V}$ obeys the symplectic constraint
\begin{equation}
\langle {\cal V}\,,\bar{\cal V}\rangle = X^I\bar F_I-F_I\bar X^I=i\,.
\end{equation}
To solve this condition, one defines
\begin{equation}
{\cal V}=e^{{\cal K}(z,\bar z)/2}v(z)\,,
\end{equation}
where $v(z)$ is a holomorphic symplectic vector,
\begin{equation}
v(z) = \left(\begin{array}{c} Z^I(z) \\ \frac{\partial}{\partial Z^I}F(Z)
\end{array}\right)\,.
\end{equation}
F is a homogeneous function of degree two, called the prepotential,
whose existence is assumed to obtain the last expression.
The K\"ahler potential is then
\begin{equation}
e^{-{\cal K}(z,\bar z)} = -i\langle v\,,\bar v\rangle\,.
\end{equation}
The matrix ${\cal N}_{IJ}$ determining the coupling between the scalars
$z^{\alpha}$ and the vectors $A^I_{\mu}$ is defined by the relations
\begin{equation}\label{defN}
F_I = {\cal N}_{IJ}X^J\,, \qquad {\cal D}_{\bar\alpha}\bar F_I = {\cal N}_{IJ}
{\cal D}_{\bar\alpha}\bar X^J\,.
\end{equation}
The bosonic action reads
\begin{eqnarray}
e^{-1}{\cal L}_{\text{bos}} &=& \frac 12R + \frac 14(\text{Im}\,
{\cal N})_{IJ}F^I_{\mu\nu}F^{J\mu\nu} - \frac 18(\text{Re}\,{\cal N})_{IJ}\,e^{-1}
\epsilon^{\mu\nu\rho\sigma}F^I_{\mu\nu}F^J_{\rho\sigma} \nonumber \\
&& -g_{\alpha\bar\beta}\partial_{\mu}z^{\alpha}\partial^{\mu}\bar z^{\bar\beta}
- V\,, \label{action}
\end{eqnarray}
with the scalar potential
\eq
V = -2g^2\xi_I\xi_J[(\text{Im}\,{\cal N})^{-1|IJ}+8\bar X^IX^J]\,,
\feq
that results from U$(1)$ Fayet-Iliopoulos gauging. Here, $g$ denotes the
gauge coupling and the $\xi_I$ are constants. In what follows, we define
$g_I=g\xi_I$.

The most general timelike supersymmetric background of the theory described
above was constructed in \cite{Cacciatori:2008ek}, and is given by
\eq
ds^2 = -4|b|^2(dt+\sigma)^2 + |b|^{-2}(dz^2+e^{2\Phi}dwd\bar w)\ , \label{gen-metr}
\feq
where the complex function $b(z,w,\bar w)$, the real function $\Phi(z,w,\bar w)$
and the one-form $\sigma=\sigma_wdw+\sigma_{\bar w}d\bar w$, together with the
symplectic section \eqref{sympl-vec}\footnote{Note that also $\sigma$ and
$\cal V$ are independent of $t$.} are determined by the equations
\eq
\partial_z\Phi = 2ig_I\left(\frac{{\bar X}^I}b-\frac{X^I}{\bar b}\right)\ ,
\label{dzPhi}
\feq
\begin{eqnarray}
&&\qquad 4\partial\bar\partial\left(\frac{X^I}{\bar b}-\frac{\bar X^I}b\right) + \partial_z\left[e^{2\Phi}\partial_z
\left(\frac{X^I}{\bar b}-\frac{\bar X^I}b\right)\right]  \label{bianchi} \\
&&-2ig_J\partial_z\left\{e^{2\Phi}\left[|b|^{-2}(\text{Im}\,{\cal N})^{-1|IJ}
+ 2\left(\frac{X^I}{\bar b}+\frac{\bar X^I}b\right)\left(\frac{X^J}{\bar b}+\frac{\bar X^J}b\right)\right]\right\}= 0\,,
\nonumber
\end{eqnarray}
\begin{eqnarray}
&&\qquad 4\partial\bar\partial\left(\frac{F_I}{\bar b}-\frac{\bar F_I}b\right) + \partial_z\left[e^{2\Phi}\partial_z
\left(\frac{F_I}{\bar b}-\frac{\bar F_I}b\right)\right] \nonumber \\
&&-2ig_J\partial_z\left\{e^{2\Phi}\left[|b|^{-2}\text{Re}\,{\cal N}_{IL}(\text{Im}\,{\cal N})^{-1|JL}
+ 2\left(\frac{F_I}{\bar b}+\frac{\bar F_I}b\right)\left(\frac{X^J}{\bar b}+\frac{\bar X^J}b\right)\right]\right\}
\nonumber \\
&&-8ig_I e^{2\Phi}\left[\langle {\cal I}\,,\partial_z {\cal I}\rangle-\frac{g_J}{|b|^2}\left(\frac{X^J}{\bar b}
+\frac{\bar X^J}b\right)\right] = 0\,, \label{maxwell}
\end{eqnarray}
\begin{equation}
2\partial\bar\partial\Phi=e^{2\Phi}\left[ig_I\partial_z\left(\frac{X^I}{\bar b}-\frac{\bar X^I}b\right)
+\frac2{|b|^2}g_Ig_J(\text{Im}\,{\cal N})^{-1|IJ}+4\left(\frac{g_I X^I}{\bar b}+\frac{g_I \bar X^I}b
\right)^2\right]\,, \label{Delta-Phi}
\end{equation}
\begin{equation}
d\sigma + 2\,\star^{(3)}\!\langle{\cal I}\,,d{\cal I}\rangle - \frac i{|b|^2}g_I\left(\frac{\bar X^I}b
+\frac{X^I}{\bar b}\right)e^{2\Phi}dw\wedge d\bar w=0\,. \label{dsigma}
\end{equation}
Here $\star^{(3)}$ is the Hodge star on the three-dimensional base with metric\footnote{Whereas
in the ungauged case, this base space is flat and thus has trivial holonomy, here we have U(1)
holonomy with torsion \cite{Cacciatori:2008ek}.}
\eq
ds_3^2 = dz^2+e^{2\Phi}dwd\bar w\ ,
\feq
and we defined $\partial=\partial_w$, $\bar\partial=\partial_{\bar w}$, as well as
\begin{equation}
{\cal I} = \text{Im}\left({\cal V}/\bar b\right)\ .
\end{equation}
Given $b$, $\Phi$, $\sigma$ and $\cal V$, the fluxes read
\begin{eqnarray}
F^I&=&2(dt+\sigma)\wedge d\left[bX^I+\bar b\bar X^I\right]+|b|^{-2}dz\wedge d\bar w
\left[\bar X^I(\bar\partial\bar b+iA_{\bar w}\bar b)+({\cal D}_{\alpha}X^I)b\bar\partial z^{\alpha}-
\right. \nonumber \\
&&\left. X^I(\bar\partial b-iA_{\bar w}b)-({\cal D}_{\bar\alpha}\bar X^I)\bar b\bar\partial\bar z^{\bar\alpha}
\right]-|b|^{-2}dz\wedge dw\left[\bar X^I(\partial\bar b+iA_w\bar b)+\right. \nonumber \\
&&\left.({\cal D}_{\alpha}X^I)b\partial z^{\alpha}-X^I(\partial b-iA_w b)-({\cal D}_{\bar\alpha}\bar X^I)
\bar b\partial\bar z^{\bar\alpha}\right]- \nonumber \\
&&\frac 12|b|^{-2}e^{2\Phi}dw\wedge d\bar w\left[\bar X^I(\partial_z\bar b+iA_z\bar b)+({\cal D}_{\alpha}
X^I)b\partial_z z^{\alpha}-X^I(\partial_z b-iA_z b)- \right.\nonumber \\
&&\left.({\cal D}_{\bar\alpha}\bar X^I)\bar b\partial_z\bar z^{\bar\alpha}-2ig_J
(\text{Im}\,{\cal N})^{-1|IJ}\right]\,. \label{fluxes}
\end{eqnarray}
In \eqref{fluxes}, $A_{\mu}$ is the gauge field of the K\"ahler U$(1)$,
\eq
A_{\mu} = -\frac i2(\partial_{\alpha}{\cal K}\partial_{\mu}z^{\alpha} -
         \partial_{\bar\alpha}{\cal K}\partial_{\mu}{\bar z}^{\bar\alpha})\,.
\feq

\section{Constant scalars}
\label{const-scal}

Let us first assume $g_I{\cal D}_{\alpha}X^I=0$, which implies that the scalars
are constant\footnote{This is true if the scalar potential has no flat directions.}.

In order to solve the system \eqref{dzPhi}-\eqref{dsigma}, inspired by the analysis
in pure gauged supergravity \cite{Cacciatori:2004rt}, we make the ansatz
\eq
\frac{\bar X}b = \frac{f(z)+p(w,\bar w)}{g(z)}\ , \qquad e^{2\Phi} = h(z)\ell(w,\bar w)\ ,
\label{ans-barXb}
\feq
where we defined $X\equiv g_IX^I$. Here, $f(z)$, $g(z)$ and $p(w,\bar w)$ are complex
functions, while $h(z)$ and $\ell(w,\bar w)$ are real. \eqref{dzPhi} implies then that
$\bar gp-g\bar p$ is independent of $w,\bar w$. This in turn leads to
\begin{displaymath}
p = (1 + i\lambda_1)\text{Re}\,p + i\lambda_2\ ,
\end{displaymath}
where $\lambda_1,\lambda_2\in\bR$ are constants. From the ansatz \eqref{ans-barXb}
it is clear that the prefactor $1+i\lambda_1$ as well as $i\lambda_2$ can be absorbed
into $f(z)$ and $g(z)$, so that we can choose $p$ real without loss of generality. But then
$g(z)$ is also real, if we want $p$ to have a nontrivial dependence on $w,\bar w$.
Thus, equ.~\eqref{dzPhi} boils down to
\eq
\partial_z\ln h = -\frac{8\text{Im}\,f}g\ , \label{Imf}
\feq
while \eqref{Delta-Phi} gives
\eq
\frac{\partial\bar\partial\ln\ell}{\ell} = h\left[-\frac14\partial^2_z\ln h + \frac2{g^2|X|^2}|f+p|^2
g_Ig_J(\text{Im}\,{\cal N})^{-1|IJ} + \frac4{g^2}(f + \bar f + 2p)^2\right]\ . \label{Delta-lnl}
\feq
This is of the type
\eq
A(w,\bar w) = -B(z) + C(z)p(w,\bar w) + D(z)p^2(w,\bar w)\ , \label{ABCD} 
\feq
for some functions $A$, $B$, $C$, $D$. Applying the operator $\partial\partial_z$ to
\eqref{ABCD} yields $p=\text{const.}$ or $C,D$ constant. The former case is trivial,
so we shall consider the latter in what follows. \eqref{ABCD} implies then that $B$ is
constant as well. Explicitely we have
\eq
D = \frac h{g^2}\left[\frac2{|X|^2}g_Ig_J(\text{Im}\,{\cal N})^{-1|IJ} + 16\right]\ , \qquad
C = (f + \bar f)D\ ,
\feq
and hence the real part of $f(z)$ is independent of $z$, and can be absorbed into $p$. One can
thus choose $f$ imaginary and $C=0$ without loss of generality. Using the special geometry relation
\eq
g^{\alpha\bar\beta}{\cal D}_{\alpha}X^I{\cal D}_{\bar\beta}
\bar X^J = -\frac 12(\text{Im}\,{\cal N})^{-1|IJ}-\bar X^I X^J\,, \label{spec-geom-rel}
\feq
we get $D=12h/g^2$. Taking into account \eqref{Imf}, the expression for $B$ becomes
\eq
B = \frac h4(\ln h)'' + \frac h{16}{(\ln h)'}^2 = \text{constant}\ . \label{expr-B}
\feq
This is a differential equation for $h$, with solution
\eq
h = \left\{\begin{array}{c@{\quad,\quad}l}
\left(\frac B{u_0} + u_0z^2\right)^2 & u_0 \neq 0\ , \\
-4Bz^2 & u_0 = 0\ , \end{array}\right.
\feq
where $u_0$ denotes a real integration constant\footnote{A further integration constant can be
eliminated by shifting $z$.}. In the following, we are interested in the case $u_0\neq 0$ only.
For the functions $g$ and $f$ one has
\eq
g = \pm 2\sqrt{\frac{3h}D}\ , \qquad f = \mp\frac i2\left(\sqrt{\frac{3h}D}\right)'\ .
\feq
By rescaling $p\to\pm p\sqrt{3/D}/2$ in the ansatz \eqref{ans-barXb} we can choose the upper
sign and set $\sqrt{3/D}/2=1$, i.e., $D=3/4$ without loss of generality. Then \eqref{Delta-lnl}
reduces to
\eq
\partial\bar\partial\ln\ell = \ell\left[-B + \frac34p^2\right]\ . \label{Delta-lnl'}
\feq
The Bianchi identities \eqref{bianchi} are automatically satisfied, while the Maxwell equations
\eqref{maxwell} imply
\eq
\partial\bar\partial p = \ell\left[\frac14p^3 - Bp\right]\ . \label{Delta-p}
\feq
As was noticed in \cite{Cacciatori:2004rt}, \eqref{Delta-lnl'} and \eqref{Delta-p} follow from
the dimensionally reduced gravitational Chern-Simons action \cite{Guralnik:2003we}
\eq
S = \int d^2x\sqrt g\left[pR + p^3\right]\ , \label{CS}
\feq
if we choose the conformal gauge $g_{ij}dx^idx^j=\ell dwd\bar w$. Note that in \eqref{CS},
$p$ is not a fundamental field, rather it is the curl of a vector potential,
$\sqrt g\epsilon_{ij}p=\partial_iA_j-\partial_jA_i$. Actually, the equations of motion following from the
action \eqref{CS} are slightly stronger than our system \eqref{Delta-lnl'}, \eqref{Delta-p}, which does
not include the traceless part of the constraints $\delta S/\delta g^{ij}=0$. Grumiller and Kummer
were able to write down the most general solution of \eqref{CS}, using the fact that the dimensionally
reduced Chern-Simons theory can be written as a Poisson-sigma model with four-dimensional
target space and degenerate Poisson tensor of rank two \cite{Grumiller:2003ad}. This solution
is given by \cite{Grumiller:2003ad}
\begin{eqnarray}
\ell &=& \frac{1 + \delta}{\cosh^4(\sqrt B\tilde x)}\ , \qquad
p = 2\sqrt B\tanh(\sqrt B\tilde x)\ , \nonumber \\
\delta &=& \left(\frac{\cal C}{2B^2}-1\right)\cosh^4(\sqrt B\tilde x)\ , \label{sol-Grumiller}
\end{eqnarray}
where the coordinate $\tilde x$ is related to $x=(w+\bar w)/2$ by
\eq
\frac{dx}{d\tilde x} = \frac{\cosh^2(\sqrt B\tilde x)}{1 + \delta}\ ,
\feq
and $\cal C$ denotes an integration constant\footnote{More precisely, $\cal C$ and $B$ are the
Casimir functions of the Poisson sigma model that can be interpreted respectively as energy
and charge \cite{Grumiller:2003ad}.}. Note that the solution for negative $B$ can be obtained
by a simple analytical continuation from \eqref{sol-Grumiller} \cite{Cacciatori:2004rt}.

The shift vector $\sigma$ can now be determined from \eqref{dsigma}, with the result
\eq
\sigma = \sigma_y dy\ , \qquad \sigma_y = \frac1{4g|X|^2}\left(\frac{p^4}{8B} - p^2
+ \frac{\cal C}B\right) - \frac{u_0p^2}{32B|X|^2}\ , \qquad y = \frac{w-\bar w}{2i}\ .
\feq
Putting all together and using $p$ as a new coordinate in place of $x$, the metric
\eqref{gen-metr} becomes\footnote{Notice that the analogue of \eqref{prePD} in minimal
gauged supergravity was found in \cite{Cacciatori:2004rt}.}
\begin{eqnarray}
ds^2 &=& -\frac{64|X|^2(B/u_0+u_0z^2)^2}{4u_0^2z^2+p^2}\left(dt + \sigma_y dy\right)^2 +
\frac{4u_0^2z^2+p^2}{16|X|^2(B/u_0+u_0z^2)^2}dz^2 \nonumber \\
&& + \frac{4u_0^2z^2+p^2}{32B|X|^2\left(\frac{p^4}{8B}-p^2+\frac{\cal C}B\right)}dp^2 +
\frac{4u_0^2z^2+p^2}{32B|X|^2}\left(\frac{p^4}{8B}-p^2+\frac{\cal C}B\right)dy^2\ . \label{prePD}
\end{eqnarray}
This resembles the Plebanski-Demianski (PD) solution \cite{Plebanski:1976gy} of cosmological
Einstein-Maxwell theory, which is also specified by quartic structure functions. Indeed,
in the case $1-{\cal C}/(2B^2)>0$, consider the coordinate transformation
\eq
\left(\begin{array}{c} \tau \\ \varsigma \end{array}\right) = \left(1 - \frac{\cal C}{2B^2}\right)^{-1/2}
\left(\begin{array}{cc} \frac1{u_0} & -\frac{\cal C}{16B^2|X|^2} \\ \frac1{4Bu_0} &
-\frac1{32B|X|^2} \end{array}\right)\left(\begin{array}{c} t \\ y \end{array}\right)\ , \qquad
q = 2u_0z\ ,
\feq
which casts the line element \eqref{prePD} into the PD form
\eq
ds^2 = \frac{p^2+q^2}{\cal P}dp^2 + \frac{\cal P}{p^2+q^2}\left(d\tau + q^2d\varsigma\right)^2
+ \frac{p^2+q^2}{\cal Q}dq^2 - \frac{\cal Q}{p^2+q^2}\left(d\tau - p^2d\varsigma\right)^2\ , \label{PD-const}
\feq
with the structure functions
\eq
{\cal P} = \gamma - \mathsf{E} p^2 + \frac{p^4}{l^2}\ , \qquad
{\cal Q} = \hat\gamma + \mathsf{E} q^2 + \frac{q^4}{l^2}\ , \label{struc-func}
\feq
where
\eq
\gamma = 32{\cal C}|X|^2\ , \qquad \hat\gamma = 64B^2|X|^2\ , \qquad
\mathsf{E} = 32B|X|^2\ , \qquad l^2 = \frac1{4|X|^2}\ . \label{PDconstants}
\feq
$l$ is related to the effective cosmological constant by $\Lambda=-3/l^2$.
In these coordinates, the fluxes \eqref{fluxes} read\footnote{Observe that contraction of
\eqref{spec-geom-rel} with $g_I$ and taking into account $g_I{\cal D}_{\alpha}X^I=0$ yields
$\bar X X^J=-\frac12(\text{Im}\,{\cal N})^{-1|IJ}g_I$, and thus $\bar X X^J$ is real.}
\eq
F^I = \frac{2X^I\bar X}{|X|(p^2+q^2)^2}(\hat\gamma - \gamma)^{1/2}\left[(p^2-q^2)(d\tau
+ q^2d\varsigma)\wedge dp + 2pq(d\tau - p^2d\varsigma)\wedge dq\right]\ .
\label{fluxesPD}
\feq
Comparing the functions \eqref{struc-func} and the field strengths \eqref{fluxesPD} with the
general expressions given in \cite{Plebanski:1976gy}, we see that in our case the mass,
nut and electric charge parameters $\mathsf{M}$, $\mathsf{N}$ and $\mathsf{Q}$ vanish.
It would be very interesting to see how one has to generalize the ansatz \eqref{ans-barXb} in
order to get solutions with nonzero $\mathsf{M}$, $\mathsf{N}$ and $\mathsf{Q}$. That there
must be supersymmetric solutions of this type is clear from the analysis for minimal gauged
supergravity in \cite{AlonsoAlberca:2000cs}.
Notice also that for $\mathsf{M}=\mathsf{N}=\mathsf{Q}=0$, the BPS conditions obtained in
\cite{AlonsoAlberca:2000cs} boil down to
\eq
{\mathsf E}^2 = \frac4{l^2}\hat\gamma\ , \label{E^2}
\feq
which is exactly what follows from \eqref{PDconstants}. As a by-product, we have thus shown
that the PD solution in minimal gauged supergravity with $\mathsf{M}=\mathsf{N}=\mathsf{Q}=0$
satisfying \eqref{E^2} does really admit a Killing spinor. This was not obvious, since
\cite{AlonsoAlberca:2000cs} analyzes only the first integrability conditions, which are in general
necessary but not sufficient for the existence of Killing spinors.

\section{Nonconstant scalar fields}
\label{nonconst-scal}

In this section we shall obtain supersymmetric rotating black holes as well as their near-horizon
geometries, which both have nontrivial moduli turned on. This is done for the SU(1,1)/U(1) model with prepotential $F=-iX^0X^1$, that has $n_V=1$ (one vector multiplet), and thus just one complex
scalar $\tau$. Choosing $Z^0=1$, $Z^1=\tau$, the symplectic vector $v$ becomes
\eq
v = \left(\begin{array}{c} 1 \\ \tau \\ -i\tau \\ -i\end{array}\right)\ .
\label{v-X0X1}
\feq
The K\"ahler potential, metric and kinetic matrix for the vectors are given
respectively by
\eq
e^{-{\cal K}} = 2(\tau + \bar\tau)\ , \qquad g_{\tau\bar\tau} = \partial_\tau\partial_{\bar\tau}
{\cal K} = (\tau + \bar\tau)^{-2}\ ,
\feq
\eq
{\cal N} = \left(\begin{array}{cc} -i\tau & 0 \\ 0 & -\frac i\tau\end{array}\right)\ .
\feq
Note that positivity of the kinetic terms in the action requires
${\mathrm{Re}}\tau>0$. For the scalar potential one obtains
\eq
V = -\frac4{\tau+\bar\tau}(g_0^2 + 2g_0g_1\tau + 2g_0g_1\bar\tau
+ g_1^2\tau\bar\tau)\ , \label{pot_su11}
\feq
which has an extremum at $\tau=\bar\tau=|g_0/g_1|$. In what follows we assume
$g_I>0$.

\subsection{1/2 BPS near-horizon geometries}
\label{1/2BPS}

An interesting class of half-supersymmetric backgrounds was obtained in \cite{Klemm:2010mc}.
It includes the near-horizon geometry of extremal rotating black holes. The metric and the
fluxes read respectively
\begin{eqnarray}
ds^2 &=& -z^2e^{\xi}\left[dt+4(e^{-2\xi}-L)\frac{dx}z\right]^2 + 4e^{-\xi}\frac{dz^2}{z^2}
\nonumber \\
&& \qquad +16e^{-\xi}(e^{-2\xi}-L)dx^2 + \frac{4e^{-2\xi}d\xi^2}{Y^2(e^{-\xi}-Le^{\xi})}\ ,
\label{metr-Y}
\end{eqnarray}
\begin{eqnarray}
F^I&=&8i\left(\frac{\bar X X^I}{1-iY}-\frac{X \bar X^I}{1+iY}\right)dt\wedge dz \\
&&+\frac 4Y\left[\frac{2\bar X X^I}{1-iY}+\frac{2X\bar X^I}{1+iY}+\left(\mbox{Im}\,\mathcal{N}
\right)^{-1|IJ}g_J\right](zdt-4Ldx)\wedge d\xi\ , \nonumber
\end{eqnarray}
where $L$ is a real integration constant and $Y$ is defined by
\eq
Y^2=64e^{-\xi}|X|^2-1\ . \label{Y}
\feq
The moduli fields $z^{\alpha}$ depend on the coordinate $\xi$ only, and obey the flow equation
\eq
\frac{dz^{\alpha}}{d\xi} = \frac i{2\bar X Y}(1-iY)g^{\alpha\bar\beta}{\cal D}_{\bar\beta}\bar X\ .
\label{dzdxi}
\feq
For $L>0$, the line element \eqref{metr-Y} can be cast into the simple form
\begin{eqnarray}
ds^2 &=& 4e^{-\xi}\left(-z^2d{\hat t}^2 + \frac{dz^2}{z^2}\right) + 16L(e^{-\xi}-Le^{\xi})
\left(dx - \frac z{2\sqrt L}d\hat t\right)^2 \nonumber \\
&& \qquad + \frac{4e^{-2\xi}d\xi^2}{Y^2(e^{-\xi}-Le^{\xi})}\ , \label{near-hor}
\end{eqnarray}
where $\hat t\equiv t/(2\sqrt L)$. \eqref{near-hor} is of the form (3.3) of
\cite{Astefanesei:2006dd}, and describes the near-horizon geometry of extremal
rotating black holes\footnote{Metrics of the type \eqref{near-hor} were discussed for the first
time in \cite{Bardeen:1999px} in the context of the extremal Kerr throat geometry.},
with isometry group $\text{SL}(2,\bR)\times\text{U}(1)$.
From \eqref{dzdxi} it is clear that the scalar fields have
a nontrivial dependence on the horizon coordinate $\xi$ unless $g_I{\cal D}_{\alpha}X^I=0$.
As was shown in \cite{Klemm:2010mc}, the solution with constant scalars is the near-horizon
limit of the supersymmetric rotating hyperbolic black holes in minimal gauged
supergravity \cite{Caldarelli:1998hg}. We shall now give an explicit example of a near-horizon
geometry with varying scalars, taking the simple model introduced above, with prepotential
$F=-iX^0X^1$. In this case the flow equation \eqref{dzdxi} becomes
\eq
\frac{d\tau}{d\xi} = \frac i{2Y}(1 - iY)\frac{-g_0+g_1\tau}{g_0+g_1\bar\tau}(\tau+\bar\tau)\ .
\feq
Using $Y$ in place of $\xi$ as a new variable, this boils down to
\eq
\frac{d\tau}{dY} = -\frac{g_1^2\tau^2-g_0^2}{2g_0g_1(Y-i)}\ ,
\feq
which is solved by
\eq
\tau = \frac{g_0}{g_1}\frac{Y-i+C}{Y-i-C}\ , \label{scal-1/2BPS}
\feq
with $C\in\bC$ an integration constant. This allows to compute $|X|^2$ as a function of $Y$,
\eq
|X|^2 = g_0g_1\frac{Y^2 + 1}{Y^2 + 1 - |C|^2}\ .
\feq
Plugging this into \eqref{Y} yields an expression for $\xi$ in terms of $Y$,
\eq
e^{-\xi} = \frac{Y^2 + 1 - |C|^2}{64g_0g_1}\ . \label{Yxi}
\feq

\subsection{Supersymmetric rotating black holes }
\label{susy-rot-bh}

We now want to obtain stationary BPS black holes with nonconstant moduli, that
approach the geometries of the previous subsection in the near-horizon limit.
To this end, we use the ansatz
\eq
\frac{\bar X^I}b = \frac{f^I(z)+\eta^I(w,\bar w)}{g(z)}\ , \qquad e^{2\Phi} = h(z)\ell(w,\bar w)\ ,
\label{ans-barXIb}
\feq
where $f^I(z)$ is an imaginary function, while $g(z)$, $\eta^I(w,\bar w)$, $h(z)$ and
$\ell(w,\bar w)$ are real. Then, \eqref{dzPhi} reduces again to \eqref{Imf}, where $f$ is
defined by $f\equiv f^Ig_I$. \eqref{Delta-Phi} becomes
\eq
\frac{\partial\bar\partial\ln\ell}{\ell} = h\left[-\frac14\partial^2_z\ln h - \frac8{g^2}\sum_I
g_I^2({\eta^I}^2 - {f^I}^2) + \frac{16}{g^2}\eta^2\right]\ , \label{ddquerlnl}
\feq
with $\eta\equiv\eta^Ig_I$. Guided by the constant scalar case (cf.~section \ref{const-scal})
we take $h/g^2=\text{const.}\equiv c_1>0$ and\footnote{It is easy to show that for constant
scalars one must have $f^I=\gamma^If$, $\eta^I=\gamma^I\eta$, where the constants
$\gamma^I$ satisfy $\gamma^Ig_I=1$. Using this together with \eqref{Imf}, equ.~\eqref{c2c1}
reduces to \eqref{expr-B}.}
\eq
-\frac h4\partial^2_z\ln h + \frac{8h}{g^2}\sum_Ig_I^2{f^I}^2 = \text{const.} \equiv c_2c_1\ .
\label{c2c1}
\feq
With these assumptions, \eqref{ddquerlnl} gives
\eq
\frac{\partial\bar\partial\ln\ell}{\ell} = c_1c_2 - 8c_1\sum_I g_I^2{\eta^I}^2 + 16c_1\eta^2\ .
\label{ddquerlnl'}
\feq
In order to solve \eqref{c2c1}, we make the ansatz
\eq
g = c+az^2\ , \qquad h = c_1(c+az^2)^2\ , \qquad f^I = i(\alpha^Iz+\beta^I)\ ,
\feq
for some real constants $a$, $c$, $\alpha^I$, $\beta^I$. One finds that \eqref{c2c1} is satisfied
if the following constraints hold:
\eq
a^2 = 8\sum_Ig_I^2{\alpha^I}^2\ , \qquad \sum_Ig_I^2\alpha^I\beta^I = 0\ , \qquad
-ac - 8\sum_Ig_I^2{\beta^I}^2 = c_2\ .
\feq
Moreover, \eqref{Imf} yields
\eq
g_I\alpha^I = -\frac a2\ , \qquad g_I\beta^I = 0\ . \label{g-alpha}
\feq
The Bianchi identities \eqref{bianchi} lead to
\eq
\alpha^I = -\frac a{4g_I}\ ,
\feq
which implies the first equation of \eqref{g-alpha}. Finally, the Maxwell equations
\eqref{maxwell} hold provided that
\eq
\alpha_{IJ}\partial\bar\partial\eta^J + 4 g_Ic_1\ell\eta\alpha_{LJ}\left(\frac ca\alpha^L\alpha^J
- \beta^L\beta^J - \eta^L\eta^J\right) = 0\ , \label{Delta-eta}
\feq
where
\eq
(\alpha_{IJ}) \equiv \left(\begin{array}{cc} 0 & 1 \\ 1 & 0\end{array}\right)\ .
\feq
It would be interesting to see if \eqref{ddquerlnl'} and \eqref{Delta-eta}, similar to \eqref{Delta-lnl'}
and \eqref{Delta-p}, follow from an action principle of the type \eqref{CS}.

We shall now solve the eqns.~\eqref{ddquerlnl'}, \eqref{Delta-eta} under the additional assumption
$\ell=\ell(x)$, $\eta^I=\eta^I(x)$, using an ansatz analogous to \eqref{sol-Grumiller}:
\begin{eqnarray}
\ell &=& \frac{1 + \delta}{\cosh^4(\kappa\tilde x)}\ , \qquad
\eta^I = \hat\eta^I\tanh(\kappa\tilde x)\ , \nonumber \\
\delta &=& A\cosh^4(\kappa\tilde x)\ , \qquad
\frac{dx}{d\tilde x} = \frac{\cosh^2(\kappa\tilde x)}{1 + \delta}\ , \label{sol-Grum'}
\end{eqnarray}
where $\kappa$, $\hat\eta^I$ and $A$ are constants. Plugging this into \eqref{ddquerlnl'} and
\eqref{Delta-eta} gives
\eq
g_0\hat\eta^0 = g_1\hat\eta^1\ , \qquad (4g_0\hat\eta^0)^2c_1 = \kappa^2 = -c_1c_2\ .
\feq
At the end, the shift vector $\sigma$ is determined by \eqref{dsigma}, which yields
\eq
\sigma = \sigma_ydy\ , \qquad \sigma_y = \frac{\hat\eta^0\kappa(\cosh^{-4}(\kappa\tilde x) + A)}
{2g_1(c+az^2)} + \frac{c_1a\hat\eta^0}{2g_1\kappa}\tanh^2(\kappa\tilde x)\ .
\feq
Similar to the case of constant scalars, for $A<0$ the solution can be cast into a
Plebanski-Demianski-type form by the coordinate transformation
\eq
\left(\begin{array}{c} t \\ y \end{array}\right) \mapsto \frac{lg_1\sqrt{\mathsf{E}}}{\hat\eta^0\sqrt{-2A}}
\left(\begin{array}{cc} \frac{al^2}2 & -\frac{(1+A)al^4\mathsf{E}}4 \\
-\frac1{\sqrt{c_1}} & \frac{l^2\mathsf{E}}{2\sqrt{c_1}} \end{array}\right)
\left(\begin{array}{c} t \\ y \end{array}\right)\ ,
\feq
\eq
p = l\sqrt{\frac{\mathsf{E}}2}\tanh(\kappa\tilde x)\ , \qquad q = \frac{al\sqrt{\mathsf{E}}}
{4\sqrt2g_0\hat\eta^0}z\ ,
\feq
where
\eq
l^2 \equiv \frac1{4g_0g_1}\ .
\feq
The metric becomes then
\begin{eqnarray}
ds^2 = && \frac{p^2+q^2-\Delta^2}{\cal P}dp^2 + \frac{\cal P}{p^2+q^2-\Delta^2}\left(dt + (q^2-\Delta^2)
dy\right)^2 \nonumber \\
&& + \frac{p^2+q^2-\Delta^2}{\cal Q}dq^2 - \frac{\cal Q}{p^2+q^2-\Delta^2}\left(dt - p^2dy\right)^2\ ,
\label{metr-PD-scal}
\end{eqnarray}
with the structure functions
\eq
{\cal P} = (1+A)\frac{\mathsf{E}^2l^2}4 - \mathsf{E}p^2 + \frac{p^4}{l^2}\ , \qquad
{\cal Q} = \frac1{l^2}\left(q^2 + \frac{\mathsf{E}l^2}2 - \Delta^2\right)^2\ ,
\feq
and the parameter $\Delta$ is defined by
\eq
\Delta \equiv \frac{\beta^0l\sqrt{\mathsf{E}}}{\sqrt2\hat\eta^0}\ .
\feq
Notice that, although we must have obviously $\mathsf{E}>0$ in the above coordinate
transformation, the final solution in the PD form can be safely continued to
$\mathsf{E}\le 0$\footnote{In fact, it is easy to see that the case of negative $\mathsf{E}$ corresponds
to the analytical continuation $\kappa=ik$, $\hat\eta^I=i\hat n^I$, where $k,\hat n^I\in\bR$. The
hyperbolic functions in \eqref{sol-Grum'} become then trigonometric.}.

In the new coordinates, the complex scalar field $\tau$ reads
\eq
\tau = \frac{g_0}{g_1}\frac{p^2 + q^2 - \Delta^2 + 2ip\Delta}{p^2 + (q-\Delta)^2}\ .
\label{compl-scal}
\feq
For $\Delta=0$, $\tau$ is thus constant, and assumes the value $\tau=g_0/g_1$, for which
the potential \eqref{pot_su11} is extremized. Note also that for $p$ fixed and $q\to\infty$
or viceversa, $\tau$ tends to $g_0/g_1$ as well. The positivity domain ${\mathrm{Re}}\tau>0$
is determined by $p^2 + q^2 - \Delta^2 > 0$. Finally, the fluxes \eqref{fluxes} are given by
$F^I=dA^I$, where
\eq
A^I = -\frac{\mathsf{E}p\sqrt{-A}}{4g_I(p^2 + q^2 -\Delta^2)}\left(dt + (q^2 - \Delta^2)dy\right)\ .
\label{A^I}
\feq
The solution is thus specified by three free parameters $A,\mathsf{E},\Delta$. A particular case is
obtained by choosing
\eq
\sqrt{-A} = \frac{l^2+j^2}{l^2-j^2}\ , \qquad \mathsf{E} = \frac{j^2}{l^2} - 1\ ,
\feq
\eq
p = j\cosh\theta\ , \qquad y = -\frac{\phi}{j\Xi}\ , \qquad t = \frac{T - j\phi}{\Xi}\ , \qquad
\Xi \equiv 1 + \frac{j^2}{l^2}\ .
\feq
Defining also
\begin{displaymath}
\rho^2 = q^2 + j^2\cosh^2\theta\ , \qquad \Delta_q = \frac1{l^2}\left(q^2 + \frac{j^2-l^2}2
- \Delta^2\right)^2\ , \qquad \Delta_{\theta} = 1 + \frac{j^2}{l^2}\cosh^2\theta\ ,
\end{displaymath}
the metric \eqref{metr-PD-scal}, scalar field \eqref{compl-scal} and U$(1)$ gauge
potentials \eqref{A^I} become
\begin{eqnarray}
ds^2 = && \frac{\rho^2-\Delta^2}{\Delta_q}dq^2 + \frac{\rho^2-\Delta^2}{\Delta_{\theta}}d\theta^2
+ \frac{\Delta_{\theta}\sinh^2\theta}{(\rho^2-\Delta^2)\Xi^2}\left(jdT - (q^2 + j^2 - \Delta^2)d\phi\right)^2
\nonumber \\
&&- \frac{\Delta_q}{(\rho^2-\Delta^2)\Xi^2}\left(dT + j\sinh^2\theta d\phi\right)^2\ , \label{metr-hyperb}
\end{eqnarray}
\eq
\tau = \frac{g_0}{g_1}\frac{j^2\cosh^2\theta + q^2 - \Delta^2 + 2ij\Delta\cosh\theta}
{j^2\cosh^2\theta + (q-\Delta)^2}\ , \label{scal-hyperb}
\feq
\eq
A^I = \frac{\cosh\theta}{4g_I(\rho^2 - \Delta^2)}\left(jdT - (q^2 + j^2 - \Delta^2)d\phi\right)\ .
\label{A^I-hyperb}
\feq
This solution contains two arbitrary constants $j$ and $\Delta$. The former can be interpreted as
rotation parameter, since for $j=0$ the geometry is static. Moreover, there is an event horizon
determined by $\Delta_q=0$, i.e., for
\eq
q^2 = q_{\text h}^2 = \Delta^2 + \frac12(l^2-j^2)\ .
\feq
From \eqref{A^I-hyperb} it is also clear that these rotating black holes carry two magnetic
charges that are inversely proportional to the coupling constants $g_I$.
Notice that the positivity domain of the scalar is $q^2 + j^2\cosh^2\theta > \Delta^2$, but since
for $q\ge q_{\text h}$ we have $q^2 + j^2\cosh^2\theta\ge q_{\text h}^2 + j^2 = \Delta^2 +
(l^2+j^2)/2 > \Delta^2$, there are no ghosts outside the horizon\footnote{Presumably there is
a curvature singularity at $q^2+j^2\cosh^2\theta = \Delta^2$, although we did not check this
explicitely.}. For $\Delta=0$, the scalar is constant, and we recover the supersymmetric rotating black
hole with hyperbolic horizon in minimal gauged supergravity found in
\cite{Caldarelli:1998hg}\footnote{Actually, even for $\Delta=0$, the above solution slightly generalizes
the one of \cite{Caldarelli:1998hg}, in that it carries two charges instead of one.}.
For $j=0$ and $\Delta\neq 0$, \eqref{metr-hyperb}-\eqref{A^I-hyperb} boil down to
\eq
ds^2 = -l^2N^2dT^2 + \frac{dq^2}{l^2N^2} + \left(q^2 - \frac{l^2}2\sinh^2\nu\right)
(d\theta^2 + \sinh^2\theta d\phi^2)\ , \label{metr-stat}
\feq
\eq
\tau = \frac{g_0}{g_1}\frac{q - \frac l{\sqrt2}\sinh\nu}{q + \frac l{\sqrt2}\sinh\nu}\ , \qquad
A^I = -\frac{\cosh\theta}{4g_I}d\phi\ , \label{tau-A^I-stat}
\feq
where
\eq
l^2N^2 = \frac{\left(\frac{q^2}{l^2} - \frac12\cosh^2\nu\right)^2}{\frac{q^2}{l^2} - \frac12\sinh^2\nu}\ ,
\qquad \Delta \equiv -\frac l{\sqrt2}\sinh\nu\ .
\feq
This is exactly the BPS black hole found in section 3.1 of \cite{Cacciatori:2009iz}, with hyperbolic
horizon and nontrivial profile for the scalar field. It is interesting to note that for zero rotation
parameter, $\tau$ becomes real, whereas for the rotating solution it is complex, i.e., one has a
nonzero axion. A similar scenario was encountered in \cite{Chong:2004na}.

Let us now take a closer look at the near-horizon geometry of \eqref{metr-hyperb}, which is
obtained by introducing new coordinates $z,\hat t,\hat\phi$ according to
\eq
q = q_{\text h} + \epsilon q_0z\ , \qquad T = \frac{\hat t q_0}{\epsilon}\ , \qquad
\phi = \hat\phi + \Omega\frac{\hat t q_0}{\epsilon}\ ,
\feq
and then taking the limit $\epsilon\to 0$. Here, $\Omega = j/(q_{\text h}^2 + j^2 -\Delta^2)$ is
the angular velocity of the horizon, and $q_0$ is defined by
\begin{displaymath}
q_0^2 = \frac{l^4\Xi^2}{8q_{\text h}^2}\ .
\end{displaymath}
In this way one gets
\begin{eqnarray}
ds^2 = &&\frac{\rho_{\text h}^2-\Delta^2}{4q_{\text h}^2z^2}l^2dz^2 + \frac{\rho_{\text h}^2-\Delta^2}
{\Delta_{\theta}}d\theta^2 + \frac{l^4\Delta_{\theta}\sinh^2\theta}{4(\rho_{\text h}^2-\Delta^2)}
\left(d\hat\phi + \frac j{q_{\text h}}zd\hat t\right)^2 \nonumber \\
&&- \frac{\rho_{\text h}^2-\Delta^2}{4q_{\text h}^2}l^2z^2d\hat t^2\ , \label{near-hor-BH}
\end{eqnarray}
where
\begin{displaymath}
\rho_{\text h}^2 = q_{\text h}^2 + j^2\cosh^2\theta\ .
\end{displaymath}
The final coordinate transformation
\eq
e^{-\xi} = \frac{q_{\text h}^2 + j^2\cosh^2\theta - \Delta^2}{16 q_{\text h}^2}l^2\ , \qquad
x = -\frac{16q_{\text h}^3}{jl^4\Xi}\hat\phi\ ,
\feq
casts the metric \eqref{near-hor-BH} into the form \eqref{near-hor}, with the constant $L$ in
\eqref{near-hor} given by
\begin{displaymath}
L = \frac{l^8\Xi^2}{1024q_{\text h}^4}\ ,
\end{displaymath}
and we used also \eqref{Yxi}. The parameter $C$ appearing in \eqref{Yxi} turns out to be related
to $\Delta$ by $\Delta^2=q_{\text h}^2|C|^2$. The phase of $C$ is fixed by requiring that the scalar
field \eqref{scal-hyperb} coincides (after taking the limit $\epsilon\to 0$) with the expression
\eqref{scal-1/2BPS}, which leads to
\begin{displaymath}
C = -i\frac{\Delta}{q_{\text h}}\ .
\end{displaymath}
Note that there is a simple relationship between $\theta$ and the coordinate $Y$ used in
section \ref{1/2BPS}, namely
\begin{displaymath}
Y = -\frac j{q_{\text h}}\cosh\theta\ ,
\end{displaymath}
and hence $Y$ is up to a prefactor identical to the coordinate $p$ that appears in the PD
form of the metric.

In conclusion, we have found a two-parameter family \eqref{metr-hyperb}-\eqref{A^I-hyperb}
of extremal rotating black holes preserving one quarter of the supersymmetries, i.e., two real
supercharges. The solutions interpolate between AdS$_4$ at infinity and the geometry
\eqref{near-hor} near the horizon, which is 1/2 BPS. Notice also that there is a nontrivial scalar
field profile \eqref{scal-hyperb}, and for $q\to q_{\text h}$, $\tau$ does not become constant,
but still depends on the horizon coordinate $\theta$.

\subsection{Lifting to M-theory}
\label{lifting}

We now want to uplift some of the black hole solutions obtained above to M-theory, and comment
on their higher-dimensional interpretation.
To this end, let us be slightly more general, and consider the stu model of ${\cal N}=2$, $D=4$
gauged supergravity (which, as we shall see below, contains the $F=-iX^0X^1$ model used in this
section as a truncation). In the zero-axion case, i.e., for real scalars, this can be embedded
into $D=11$ supergravity using the reduction ansatz presented in \cite{Cvetic:1999xp},
that we briefly review in what follows. The eleven-dimensional metric reads
\eq
ds_{11}^2 = \tilde\Delta^{2/3}ds_4^2 + g^{-2}\tilde{\Delta}^{-1/3}\sum_{I=0}^3{X^I}^{-1}
\left(d\mu_I^2 + \mu_I^2(d\phi_I + gA^I)^2\right)\ ,
\feq
where $\tilde\Delta=\sum_{I=0}^3X^I\mu_I^2$. The four quantities $\mu_I$ satisfy $\sum_I\mu_I^2=1$,
and can be parametrized in terms of angles on S$^3$ as
\begin{displaymath}
\mu_0 = \sin\vartheta\ , \qquad \mu_1 = \cos\vartheta\sin\chi\ , \qquad \mu_2 =
\cos\vartheta\cos\chi\sin\psi\ , \qquad \mu_3 = \cos\vartheta\cos\chi\cos\psi\ .
\end{displaymath}
The $X^I$ are given by
\eq
X^I = e^{-\frac12\vec a_I\cdot\vec\varphi}\ , \qquad \vec\varphi = (\varphi_1,\varphi_2,\varphi_3)\ ,
\label{X^I-red}
\feq
with
\begin{displaymath}
\vec a_0 = (1,1,1)\ , \qquad \vec a_1 = (1,-1,-1)\ , \qquad \vec a_2 = (-1,1,-1)\ , \qquad
\vec a_3 = (-1,-1,1)\ ,
\end{displaymath}
and satisfy $X^0X^1X^2X^3=1$.

The reduction ansatz for the four-form field strength is
\begin{eqnarray}
F_{(4)} &=& 2g\sum_I\left({X^I}^2\mu_I^2 - \tilde\Delta X^I\right)\epsilon_{(4)} + \frac1{2g}
\sum_I{X^I}^{-1}\bar\ast dX^I\wedge d(\mu_I^2)\nonumber \\
&& -\frac1{2g^2}\sum_I{X^I}^{-2}d(\mu_I^2)\wedge(d\phi_I + gA^I)\wedge\bar\ast F^I\ , \label{4-form}
\end{eqnarray}
where $F^I=dA^I$, $\bar\ast$ denotes the Hodge dual operator of $ds_4^2$ and
$\epsilon_{(4)}$ is the corresponding volume form.

This leads to the four-dimensional stu model with bosonic action
\eq
e^{-1}{\cal L}_4 = \frac12\left[R - \frac12\left(\partial\vec\varphi\right)^2 - \frac14\sum_I
e^{\vec a_I\cdot\vec\varphi}{F^I}^2 + 8g^2(\cosh\varphi_1 + \cosh\varphi_2 + \cosh\varphi_3)\right]\ ,
\label{stu}
\feq
which can also be obtained from the general theory \eqref{action} by choosing a prepotential
proportional to $(X^0X^1X^2X^3)^{1/2}$, taking all $g_I$ equal, and
subsequently setting the axions to zero, cf.~\cite{Cacciatori:2009iz} for details\footnote{The $X^I$
used here are related to the $X^I$ in section 3.2 of \cite{Cacciatori:2009iz} by
$X^I_{\text{here}}=2\sqrt2 X^I_{\text{there}}$. Moreover, one has to identify $g_I=g/\sqrt2$,
$F^I_{\text{here}}=\sqrt2 F^I_{\text{there}}$, and $e^{\varphi_{\alpha}}=\tau^{\alpha}$, $\alpha=1,2,3$.}.
In order to obtain from \eqref{stu} the model with $F=-iX^0X^1$ considered in this section, one has
to further truncate according to
\eq
\varphi_1 = \varphi_3 = 0\ , \qquad e^{\varphi_2} = \tau\ , \qquad F^2 = F^0\ , \qquad F^3 = F^1\ ,
\feq
such that $X^2=X^0$ and $X^3=X^1$. This yields exactly the model introduced in
eqns.~\eqref{v-X0X1} ff., with the additional restriction that $\tau$ must be real\footnote{Notice
that $X^0, X^1$ computed from \eqref{v-X0X1} differ from $X^0$ and $X^1$ defined in \eqref{X^I-red}
by a factor one half.} and $g_0=g_1=g$.

Due to the zero-axion condition $\tau=\bar\tau$, we
cannot uplift the rotating black holes \eqref{metr-hyperb}-\eqref{A^I-hyperb}, since these have
a complex scalar unless $\Delta=0$\footnote{Actually, the $F=-iX^0X^1$ model can be
embedded into ${\cal N}=4$, $D=4$ $\text{SO}(4)$ gauged supergravity as well. To see this,
set $\tau=e^{-\varphi}+i\chi$, which casts the action \eqref{action} into an abelian truncation
of the bosonic ${\cal N}=4$ $\text{SO}(4)$ gauged supergravity action (17) of \cite{Cvetic:1999au}.
Solutions of the latter can (even for $\chi\neq 0$) in principle be lifted to eleven dimensions along
the lines of \cite{Cvetic:1999au}, but we shall leave this for future work.}.
For the static solution \eqref{metr-stat}, \eqref{tau-A^I-stat},
$\tau$ is real, and the above reduction ansatz leads to an eleven-dimensional metric
\begin{eqnarray}
ds_{11}^2 &=& \tilde\Delta^{2/3}ds_4^2 + g^{-2}\tilde\Delta^{-1/3}\left\{\tau^{1/2}\left[d\mu_0^2
+ \mu_0^2 (d\phi_0 - \frac14\cosh\theta d\phi)^2 + d\mu_2^2\right.\right. \nonumber \\
&& \left.\left.+ \mu_2^2 (d\phi_2 - \frac14\cosh\theta d\phi)^2\right] + \tau^{-1/2}\left[d\mu_1^2
+ \mu_1^2(d\phi_1 - \frac14\cosh\theta d\phi)^2\right.\right. \nonumber \\
&&\left.\left. + d\mu_3^2 + \mu_3^2(d\phi_3 - \frac14
\cosh\theta d\phi)^2\right]\right\}\ , \label{metr-11d}
\end{eqnarray}
where
\begin{displaymath}
\tilde\Delta = \tau^{-1/2}(\mu_0^2 + \mu_2^2) + \tau^{1/2}(\mu_1^2 + \mu_3^2)\ ,
\end{displaymath}
while $ds_4^2$ and $\tau$ are given by \eqref{metr-stat} and \eqref{tau-A^I-stat} (with
$g_0=g_1=g$) respectively. Finally, the four-form field strength can be easily computed
from \eqref{4-form}.

In the case $\nu=0$, the scalar $\tau$ becomes constant ($\tau=1$), and the solution
\eqref{metr-11d} can be interpreted as the gravity dual corresponding to membranes
wrapping holomorphic curves in a Calabi-Yau five-fold \cite{Gauntlett:2001qs}.
Moreover, for $\nu=0$ (i.e., $\Delta=0$), $\tau$ in \eqref{scal-hyperb} is real (and constant),
which allows to uplift also the rotating black holes, which in eleven dimensions correspond
to supersymmetric waves on wrapped membranes \cite{Gauntlett:2001qs}. It would be
interesting to see whether the general solution \eqref{metr-11d} (for $\nu\neq 0$) has a similar
interpretation. This might allow for a microscopic entropy computation of the four-dimensional
black hole \eqref{metr-stat}, which can then be compared with the macroscopic
Bekenstein-Hawking result
\eq
S_{\text{BH}} = \frac{A_{\text{hor}}}{4G_4} = \frac{\pi V}{4g^2}\ , \qquad V \equiv
\int\sinh\theta d\theta d\phi\ ,
\feq
where we used that $8\pi G=1$ in our conventions. Notice that, for a noncompact horizon H$^2$,
only the entropy density $s=S/V$ is finite. If instead the hyperbolic space is compactified to a
Riemann surface of genus $h$, we can use Gauss-Bonnet to get $V=4\pi(h-1)$, and thus
\eq
S_{\text{BH}} = \frac{\pi^2}{g^2}(h-1)\ .
\feq

\section{Final remarks}
\label{final}

In this paper we have constructed new magnetically charged rotating BPS black holes in
${\cal N}=2$, $D=4$ gauged supergravity coupled to abelian vector multiplets. One of our results
is a two-parameter family of solutions with noncompact horizon that preserve two real
supercharges and have a nontrivial
scalar field profile. In the near-horizon limit, there is a supersymmetry enhancement to 1/2 BPS.
We limited the calculations of section \ref{nonconst-scal} to the prepotential $F=-iX^0X^1$,
but it would be very interesting to generalize them to the stu or at least to the so-called t$^3$ model,
since this admits spherically symmetric static BPS black holes \cite{Cacciatori:2009iz}, that
can in principle be given rotation.

A further point to explore would be how the attractor equations in gauged
supergravity \cite{Cacciatori:2009iz,Dall'Agata:2010gj} get modified if one includes
rotation. This involves solving eqns.~\eqref{dzPhi}-\eqref{dsigma} for the most general
stationary near-horizon geometry.

Finally, another possible generalization of our work is the inclusion of hypermultiplets.
Black holes in ${\cal N}=2$, $D=4$ gauged supergravity with charged hypers were
constructed and analyzed in \cite{Hristov:2010eu}. When the hypermultiplet scalars
are charged, black holes of this type might have applications in the emerging field
of holographic superconductivity, where usually no analytical solution is known, and
one has to resort to numerical techniques.

Work along these directions is in progress \cite{Colleoni:2011}.

\acknowledgments

This work was partially supported by INFN and MIUR-PRIN contract 20075ATT78.

\end{document}